\documentclass[aps,twocolumn,amsmath,amssymb]{revtex4-1}

\usepackage{graphicx}
\newcommand{\circled}[1]{\raisebox{.9pt}{\textcircled{\raisebox{-.9pt}{#1}}}}

\begin{document}

\title{Influence of device geometry and imperfections on the interpretation of transverse magnetic focusing experiments}

\author{Yik K. Lee}
\email{yik.kheng.lee@student.rmit.edu.au}
\affiliation{Theoretical, Computational, and Quantum Physics, School of Science, RMIT University, Melbourne, Australia}
\affiliation{ARC Centre of Excellence in Future Low-Energy Electronics Technologies, RMIT University, Melbourne, Australia}

\author{Jackson S. Smith}
\affiliation{Theoretical, Computational, and Quantum Physics, School of Science, RMIT University, Melbourne, Australia}
\affiliation{ARC Centre of Excellence in Future Low-Energy Electronics Technologies, RMIT University, Melbourne, Australia}

\author{Jared H. Cole}
\email{jared.cole@rmit.edu.au}
\affiliation{Theoretical, Computational, and Quantum Physics, School of Science, RMIT University, Melbourne, Australia}
\affiliation{ARC Centre of Excellence in Future Low-Energy Electronics Technologies, RMIT University, Melbourne, Australia}

\begin{abstract} 
Spatially separating electrons of different spins and efficiently generating spin currents are crucial steps towards building practical spintronics devices.
Transverse magnetic focusing is a potential technique to accomplish both those tasks.
In a material where there is significant Rashba spin--orbit interaction, electrons of different spins will traverse different paths in the presence of an external magnetic field.
Experiments have demonstrated the viability of this technique by measuring conductance spectra that indicate the separation of spin--up and spin--down electrons.
However the effect that the geometry of the leads has on these measurements is not well understood.
By simulating an InGaAs-based transverse magnetic focusing device, we show that the resolution of features in the conductance spectra is affected by the shape, separation and width of the leads.
Furthermore, the number of subbands occupied by the electrons in the leads affects the ratio between the amplitudes of the spin--split peaks in the spectra.
We simulated devices with random onsite potentials and observed that transverse magnetic focusing devices are sensitive to disorder.
Ultimately we show that careful choice and characterisation of device geometry are crucial for correctly interpreting the results of transverse magnetic focusing experiments.
\end{abstract}

\maketitle

\section{Introduction}

Unlike conventional electronics, spintronics manipulates the spin degree of freedom of an electron to transfer and store information~\cite{Xia2012}.
One of the challenges in spintronics is to generate a spin current efficiently~\cite{Hirohata2020}, and another is to spatially separate spin currents~\cite{Lo2017}.
Transverse magnetic focusing (TMF) offers a solution to both these challenges by exploiting the spin--momentum coupling induced by the Rashba spin--orbit interaction~\cite{Rashba1984}.
Electrons injected into a two-dimensional electron gas (2DEG) with a significant Rashba spin--orbit coupling are separated by their spins and travel in different cyclotron paths as shown in Fig.~\ref{fig:TMF_system}a when a perpendicular magnetic field is applied to the device.
Measuring the conductance between the points of electron injection and collection, while varying the strength of this magnetic field, gives a spectrum like the one shown in Fig.~\ref{fig:TMF_system}b, with peaks occurring at magnetic field strengths that focus an electron at the point of collection.

TMF in a metal was first demonstrated by Tsoi in 1974 using a Bi single crystal~\cite{Tsoi1974}. 
TMF in a 2DEG was shown by van Houten \textit{et al.} in 1988 using a GaAs/AlGaAs heterostructure at temperatures ranging from 0.05~K to 7~K~\cite{VanHouten1989}.
In 1992, Heremans \textit{et al.} observed ballistic hole transport by performing a TMF experiment in a two-dimensional hole gas (2DHG) using a GaAs/AlGaAs heterostructure~\cite{Heremans1992, Heremans1994}.
More recently, TMF has been shown to be a viable spin filter in a wide range of experiments~\cite{Potok2002,Folk2003,Rokhinson2004,Dedigama2006,Heremans2007a,Rendell2015a,Lo2017,Yan2017,Yan2018, Tsoi1999, Bozhko2014}. 
There are also indications that TMF need not be limited to semiconductor heterostructures, as Taychatanapat \textit{et al.} demonstrated it can be performed in graphene devices up to a temperature of 300~K~\cite{Taychatanapat2013}.

In a TMF experiment, spatial separation of the spin currents is indicated by the presence of a pair of spin--split peaks in the conductance spectrum that each correspond to a different spin, such as the peaks at $B^{}_{{z}}= 0.061$~T and 0.089~T in Fig.~\ref{fig:TMF_system}b.
In materials without Rashba spin--orbit coupling, the spin currents would not spatially separate, resulting in a single higher peak, similar to the peak at $B^{}_{{z}}= 0.151$~T in Fig.~\ref{fig:TMF_system}b.
The spin--split peaks do not have equal amplitudes, which has previously been attributed to spin polarisation in the injected electrons~\cite{Rokhinson2004,Reynoso2007}.
The wave-like nature of electrons suggests that there should be wave interference, which can be observed as interference fringes in the spectrum~\cite{Usaj2004,Bladwell2017}.
Although transport in these devices is considered ballistic, there are many factors that could affect measurements in experiments, such as temperature~\cite{Potok2002,Heremans2007a,Yan2017} and disorder. 
The effects of physical device parameters such as the shape and width of the leads used remain unclear.
Square-shaped~\cite{Lo2017}, trumpet-shaped~\cite{Dedigama2006} and funnel-shaped~\cite{Potok2002} leads have all been used in experiments.

In this paper we focus on computing TMF measurements in an InGaAs-based 2DEG with Rashba spin--orbit coupling with realistic device geometries.
We simulate TMF using wave function matching as implemented through the program KWANT~\cite{Groth2014} using a finite difference method with a square grid discretisation scheme.
A brief introduction of TMF is given in Section~\ref{section:TMF}.
We then present our numerical approach in Section~\ref{section:Method}.
The effects that the shape, separation distance and width of the leads have on the TMF response are presented in Section~\ref{section:Shape}.
The effects of varying the number of subbands occupied by the injected electrons are presented in Section~\ref{section:Fermi}.
We simulate the leads as quantum point contacts (QPCs) in Section~\ref{section:QPCs} and show that effects seen in previous sections are linked in this more realistic model.
We investigate the influence that disorder in the surface potential of a 2DEG has on the TMF results in Section~\ref{section:Disorder} and conclude in Section~\ref{section:Conclusion}.

\section{Methods}
\subsection{Semi-classical model}
\label{section:TMF}
\begin{figure}[ht]
    \includegraphics[width=0.95\linewidth]{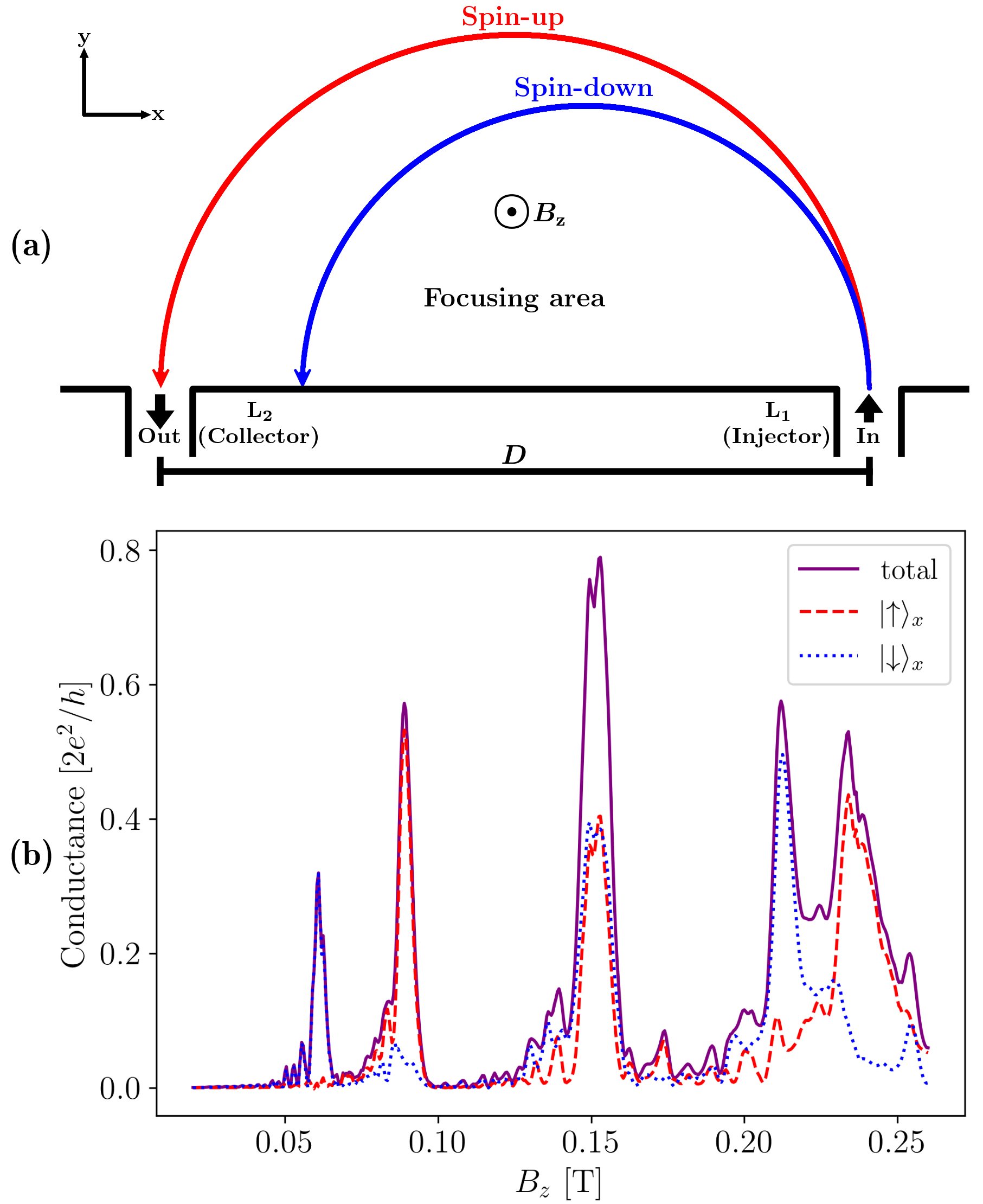}
    \caption{
    (a) Schematic of a typical transverse magnetic focusing device. 
    Electrons injected at lead $\mathrm{L}^{}_{1}$ are focused and collected at lead $\textrm{L}^{}_{2}$ by applying an out-of-plane magnetic field $B^{}_{{z}}$. 
    If the Rashba effect is present, spin--up (red) and spin--down (blue) electrons have different trajectories due to spin--orbit interaction. 
    Distance between the leads ($D$) is defined from the middle of each lead. 
    (b) Example conductance spectrum for a transverse magnetic focusing device showing conductance as a function of the out-of-plane magnetic field calculated using the method described in Section~\ref{section:Method}. 
    The solid line (purple) shows the total conductance, the dashed (red) and dotted (blue) lines show the spin projected conductance for spin--up and spin--down in $x$ respectively.
    }
    \label{fig:TMF_system}
\end{figure}

A typical TMF experiment consists of a two-dimensional focusing area connected to two leads which serve as injector and collector for electrons (see Fig.~\ref{fig:TMF_system}a).
An out-of-plane magnetic field $B^{}_{{z}}$ is applied such that an electron injected into the system experiences a Lorentz force which causes it to move in a cyclotron orbit with a Larmor radius $R^{}_{\mathrm{L}}$ given by~\cite{VanHouten1989}:
\begin{equation}
    R^{}_{\mathrm{L}}=\frac{\hbar k}{|e|B^{}_{{z}}}
    \label{Eq:larmor}
\end{equation}
where $\hbar$ is the reduced Planck constant, $k$ is the magnitude of the wavevector of the electron, and $e$ is the charge of an electron. 

At particular values of $B^{}_{{z}}$, where the distance between the leads coincides with integer multiples of $2R^{}_{\mathrm{L}}$, the probability of transmission for an electron between these two leads is nonzero as long as the reflections from the wall are specular. 
This is observed as sharp peaks in the conductance (as shown in Fig.~\ref{fig:TMF_system}b).
Experiments usually measure current and resistance, which are proportional to the conductance spectra we calculate in this paper.

When a strong Rashba spin--orbit interaction is present in the system, spin--up and spin--down electrons entering the 2DEG with the same energy have different crystal momenta. 
Consequently, spin--up and spin--down electrons have orbits with different Larmor radii, which results in the spatial separation of spin currents as indicated in Fig.~\ref{fig:TMF_system}a and the presence of spin--split peaks in the conductance spectrum as seen in Fig.~\ref{fig:TMF_system}b. 
The Rashba interaction is described by the Hamiltonian~\cite{Rashba1984}:
\begin{equation}
\label{Eq:Rashba}
    H^{}_{\mathrm{R}}=\alpha\left(\boldsymbol{\sigma} \times \boldsymbol{k}\right)\cdot \hat{z}
\end{equation}
where $\alpha$ is the Rashba coupling parameter, $\boldsymbol{\sigma}$ is the Pauli vector, $\boldsymbol{k}$ is the wavevector of the electron, and $\hat{z}$ is the unit vector normal to the plane of the 2DEG.

The dispersion relation of a 2DEG with Rashba interaction is then given by~\cite{Rashba1984}:
\begin{align}
    E^{}_{\pm} &= \frac{\left(\hbar k \right)^{2}}{2m^{*}} \pm \alpha k \nonumber\\
    E^{}_{\pm} &= \frac{1}{2m^{*}}\left[\left(\hbar k \pm \frac{m^{*}\alpha}{\hbar}\right)^{2} - \left(\frac{m^{*}\alpha}{\hbar}\right)^{2}\right]
    \label{Eq:E_Rashba}
\end{align}
where $E^{}_{\pm}$ are the energies of the spin--up($+$)/down($-$) electrons, $k=|\boldsymbol{k}|$, and $m^{*}$ is the effective mass of the electron.

The magnetic field strengths, $B^{+}_{{z}}$($B^{-}_{{z}}$), required to focus spin--up(spin--down) electrons of the same Fermi energy, $E^{}_{\mathrm{F}}$, into orbits of the same Larmor radius will be~\cite{Rokhinson2004}:
\begin{align}
    B^{\pm}_{{z}} &= \frac{1}{|e| R^{}_{\mathrm{L}}} \left( \sqrt{2 m^{*} E^{}_{\mathrm{F}}} \mp \frac{m^{*}\alpha}{\hbar} \right)
    \label{Eq:B_up_down}
\end{align}

The difference between these two magnetic field strengths is the separation between the spin--up and spin--down peak pairs in the conductance spectrum.
This semi-classical approximation ignores quantum effects such as the Berry curvature correction to the velocity of the electron~\cite{Xiao2010}.
However, for the parameter regime we consider, such corrections are several orders of magnitude smaller than the velocities.
Thus we do not consider such corrections in this paper.

It is also worth noting that classically the Larmor radius (Eq.~\ref{Eq:larmor}) is written in terms of the electron's tangential velocity.
In a semi-classical treatment, one might therefore be tempted to substitute the group velocity (${v_{\mathrm{g}} = \frac{1}{\hbar}\frac{\partial E}{\partial k}}$) for the tangential velocity, instead of substituting the crystal momentum ($\hbar k$) for its counterpart in the classical theory.
There is a subtlety in the semi-classical treatment, however, where both the Larmor radius and the cyclotron frequency depend on the spin state of the electron via the Rashba coupling, and these effects cancel such that the electron's velocity is independent of spin~\cite{Reynoso2004}.
In other words, the fact that the velocities are the same for the two spin states does not mean their Larmor radii are equal, as their cyclotron frequencies also differ.
Our numerical approach, which is fully quantum mechanical, explicitly solves for the wave propagation of the electron via the system's Hamiltonian and therefore does not depend on the definition of the Larmor radius.
Nevertheless we find our results correspond well to the radii predicted by equations~\ref{Eq:larmor} and \ref{Eq:B_up_down}.

\begin{figure}[htp]
    \includegraphics[width=0.95\linewidth]{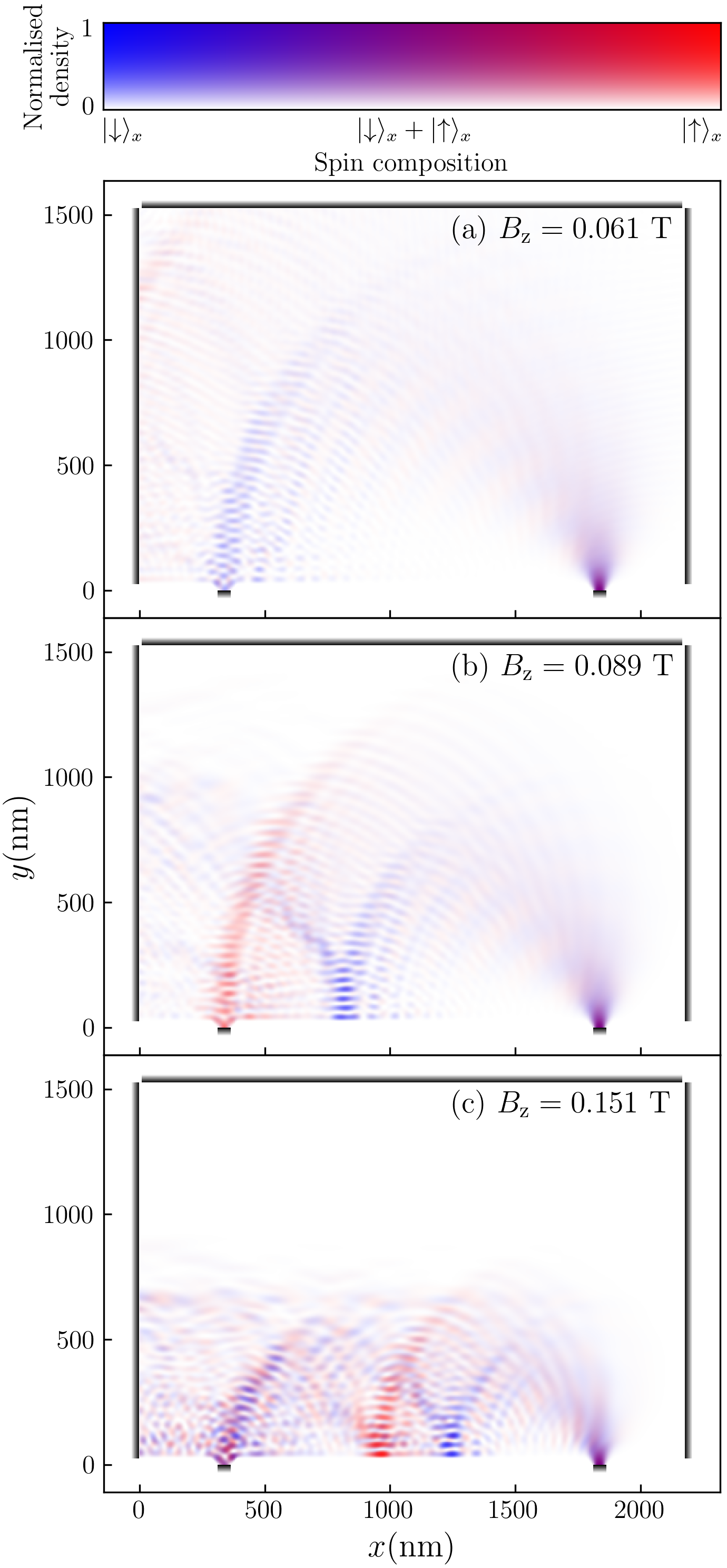}
    \caption{
    spin--projected probablity densities for an electron injected into a TMF device with an out-of-plane magnetic field of (a) 0.061~T, (b) 0.089~T, and (c) 0.151~T, corresponding to the first three peaks in Fig.~\ref{fig:TMF_system}(b).
    The colour in the plots indicate the direction of spin as measured along the $x$ axis. 
    The ratio of red to blue in the RGB colour values at each point is set to be proportional to the expected spin states at that point.
    Pure red indicates a pure spin--up state and pure blue indicates a pure spin--down state, while purple indicates a mixed state.
    The transparency values are on a normalised tanh scale, given by $\tanh(2d_{i}/d_{\mathrm{max}})/\tanh(2)$, where $d_{i}$ is the density at each point and $d_{\mathrm{max}}$ is the maximum density in each plot.
    Points with the maximum density in each plot will be completely opaque, while points with zero density will be completely transparent.
    }
    \label{fig:LDOS}
\end{figure}

We can now discuss the interpretation of a typical TMF response using this semi-classical model.
The conductance between the leads varies with magnetic field strength, as illustrated in Fig.~\ref{fig:TMF_system}b.
The first pair of peaks located at approximately 0.061~T and 0.089~T correspond to when the Larmor radii of the spin--down and spin--up electrons coincide with the distance between the leads respectively, resulting in almost pure spin currents.
The effect can be seen more clearly by plotting the probability density of conduction electrons injected into a TMF device.
In Fig.~\ref{fig:LDOS}a, b we see unfiltered spin--down and spin--up electrons spatially separate into distinct trajectories in the focusing area.

The peak at $B^{}_{{z}}\approx0.151$~T in Fig.~\ref{fig:TMF_system}b shows where the spin currents combine into a single current, giving a much higher total conductance. 
If we continue increasing the magnetic field strength, we eventually get another pair of peaks around 0.225~T that are less distinct compared to the first pair.
Our numerical results show good agreement with the semi-classical approximation of the peak positions calculated using Eq.~\ref{Eq:B_up_down}.
More sophisticated semi-classical approximations can also allow the finer features in the conductance spectrum to be analytically calculated~\cite{Bladwell2017}.

\begin{figure*}[htp]
    \centering
    \includegraphics[width=0.95\linewidth]{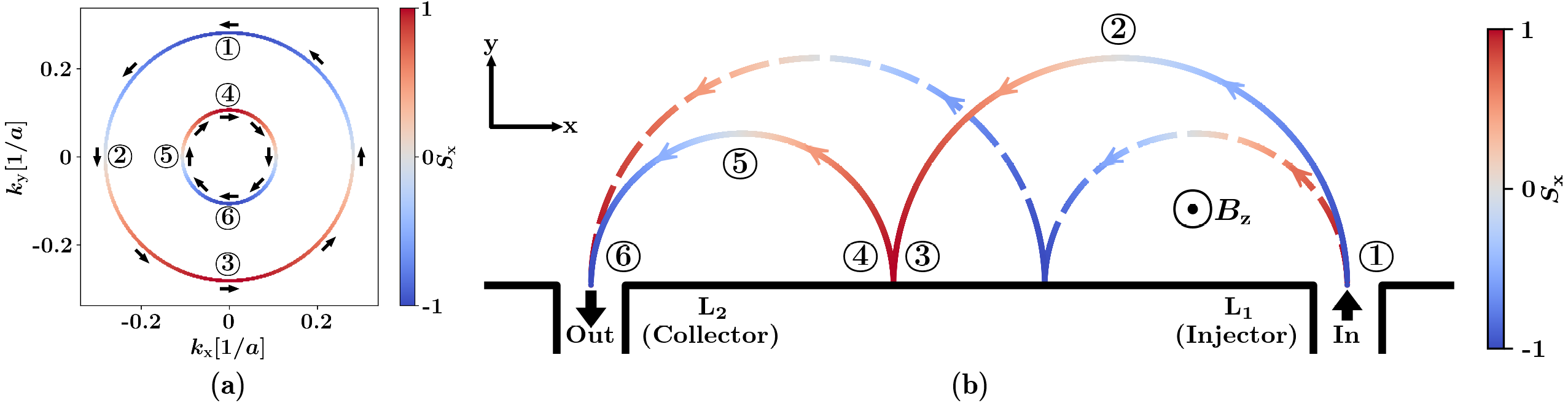}
    \caption{
    (a) Slice of the Fermi surface of a 2DEG with Rashba spin orbit coupling at a certain energy.
    The directions of the arrows indicate the spin alignment of the electron in the $xy$ plane, and the colour of the bands indicates the spin of the electron measured along the $x$ direction.
    (b) Classical trajectories of electrons injected into the scattering area of a TMF device with an out-of-plane magnetic field $B_{{z}}$ applied.
    The colour of the lines indicates the $x$-component of the spin of the electron, blue indicates spin--down, red indicates spin--up and grey indicates zero spin in $x$.
    The solid line shows the path of electrons with spin--down in $x$ at the injection point, and the dashed line shows the path of electrons with spin--up in $x$ at the injection point.
    Circled numbers in both figures label corresponding positions in real space and $k$-space.
    }
    \label{fig:2D_band}
\end{figure*}

The recombination of the spatially separated spin currents upon reflection from a wall of the device is a known phenomenon~\cite{Usaj2004,Lo2017}, but deserves more discussion as the reason it happens may not immediately be clear.
The probability density plot in Fig.~{\ref{fig:LDOS}}c corresponds to the highest peak at approximately 0.151~T in Fig.~\ref{fig:TMF_system}b.
We see that the spatially separated trajectories of the spin--up and spin--down electrons overlap at the collector after reflecting off the bottom edge of the device.
This suggests the crystal momenta of the electrons change after reflection from a hard wall boundary as the Larmor radius of each trajectory has changed.

The picture of the spin--up and spin--down electrons moving in their own trajectories as shown in Fig.~\ref{fig:TMF_system}a is greatly simplified and commonly used in the TMF literature.
In this picture, spin--up and down are defined to be parallel or anti-parallel to the Rashba effective magnetic field, which is always perpendicular to the direction of the instantaneous velocity of the electron.
In contrast, if a stationary frame is chosen for the definition of spin--up and spin--down (\textit{e.g.}~Cartesian coordinates in the lab frame), it becomes more obvious that the spin of the electron precesses as it travels through the focusing area~\cite{Reynoso2004}.
Figure~\ref{fig:2D_band}a illustrates this; the spin of the electron is always aligned tangentially to the Fermi surface in $k$-space, and at points where the electron is moving only in the $x$ or $y$ direction, its spin is aligned along the $y$ or $x$ direction, respectively.

If we follow the path of an electron that has spin--down in $x$ when injected into the focusing area as shown by the solid line in Fig.~\ref{fig:2D_band}b, we see how the spin precession leads to the spin currents recombining after reflecting off a hard wall boundary in the device.
At the injection point, labelled \circled{1}, the electron is moving in the positive $y$ direction, which corresponds to point \circled{1} on the outer Fermi surface in $k$-space.
It has a larger crystal momentum than an electron with spin--up in $x$ (which starts on the inner Fermi surface at \circled{4}) and therefore a larger Larmor radius (Eq.~\ref{Eq:larmor}).
Here we see that the group velocity is proportional to $\frac{\partial E}{\partial k}$, and the gradients of the inner and outer Fermi surfaces are equal in magnitude, so the spin--up and spin--down electrons have equal group velocities.
Although the the group velocities have the same signs as the corresponding crystal momenta at this particular energy, this is not necessarily always the case, \textit{e.g.}~when the Fermi surfaces are no longer concentric near the bottom of the band.

As the electron moves through the focusing area, it flips to spin--up in $x$ after point \circled{2}.
When it reaches the boundary of the device at point \circled{3}, the electron is moving in the negative $y$ direction before it is reflected.
After reflection, spin is conserved but velocity changes sign, \textit{i.e.}~it switches from point \circled{3} on the outer Fermi surface to point \circled{4} on the inner Fermi surface, similar to the effect observed by Chen \textit{et al.}~\cite{Chen2005}.
The electron now has a lower crystal momentum and a shorter Larmor radius as it continues moving through point \circled{5}, where it flips back to spin--down in $S^{}_{{x}}$, and exits the device at point \circled{6}.
An electron which is spin--up in $S^{}_{{x}}$ at injection would go through a similar process, but starts with a lower $k$ at injection and changes to a higher $k$ after reflection.
Thus the separate spin currents recombine after reflecting off the boundary of the system once (as seen in Fig.~\ref{fig:LDOS}c), resulting in a sharp peak in the conductance spectrum.

\subsection{Numerical model}
\label{section:Method}
For all calculations in this paper, we simulated a TMF device with a geometry as shown in Fig.~\ref{fig:TMF_system}a.
The injection and collection leads of the device were separated by 1500~nm as measured from the centre of each lead.
The left, right and top edges of the device are simulated as open boundaries by attaching semi-infinite leads to them.
The bottom edge is a hard wall boundary except for the injection and collection leads.
An additional 312.5~nm was added between the side edges and the leads to serve as a buffer.
The size of this buffer was chosen to be large enough that it did not significantly affect the results.
The height of the device (length in $y$ direction) was chosen to be the equal to the length of the separation between the contacts, 1500~nm, so that the focusing area was large enough to accommodate the cyclotron trajectory of an electron focused from the injection lead to the collection lead.

The device was described as a 2DEG confined to the $xy$ plane, with an external magnetic field $B^{}_{{z}}$ applied along the $z$ axis and Rashba spin--orbit coupling with the following Hamiltonian~\cite{Usaj2004}:
\begin{equation}
    \begin{split}
    H = -\frac{\hbar^2}{2m^{*}}\left( \frac{\partial^2}{\partial x^2} + \frac{\partial^2}{\partial y^2} \right) - \frac{g}{2} \mu^{}_{\mathrm{B}} \sigma^{}_{{z}} B^{}_{{z}} \\
    - i \alpha \left( \frac{\partial}{\partial y} \sigma^{}_{{x}} - \frac{\partial}{\partial x} \sigma^{}_{{y}} \right)
    \end{split}
    \label{Eq:Hamiltonian}
\end{equation}
where $g$ is the effective g-factor, $\mu^{}_{B}$ is the Bohr magneton, and $\sigma^{}_{x,y,z}$ are the Pauli spin matrices.
To approximate the effect of the magnetic field on the electron we used the Peierls substitution~\cite{Peierls1933a}:
\begin{equation}
    -i\frac{\partial}{\partial x} \Rightarrow -i \frac{\partial}{\partial x} + \frac{e}{\hbar} A^{}_{x}
    \label{Eq:Peierls}
\end{equation}
where $A^{}_{x}$ is the $x$-component of the magnetic vector potential.
The $y$-components were similarly substituted.
The Landau gauge in $\hat{y}$, $A=-xB^{}_{{z}}\hat{y}$, was used everywhere except in the semi-infinite leads on the left and right edges and a thin strip of the focusing area next to those leads, where the Landau gauge in $\hat{x}$, $A=-yB^{}_{{z}}\hat{x}$, was used instead.
We used the effective mass $m^{*} = 0.04m^{}_{e}$ where $m^{}_{e}$ is the free electron mass, and an effective g-factor $g = 9$ for InGaAs~\cite{Chuang2015}. 
For the Rashba coupling parameter, we used the value for an InGaAs-based 2DEG, $\alpha=3.1 \times 10^{-11}$~eV$\cdot$m~\cite{Lo2017}.
We note that other spin--orbit interactions such as the Dresselhaus effect are present in InGaAs~\cite{Kawaguchi2019}, however they are not as strong as the Rashba effect, therefore we only included the Rashba effect in this paper for simplicity.
The external magnetic field and Rashba spin--orbit coupling were applied everywhere in the system, including within the leads.
Note that while we refer to the injection and collection points as \textit{leads}, they are still part of the 2DEG, and typically correspond to QPCs in experiments.

\begin{figure}[ht]
    \includegraphics[width=0.95\linewidth]{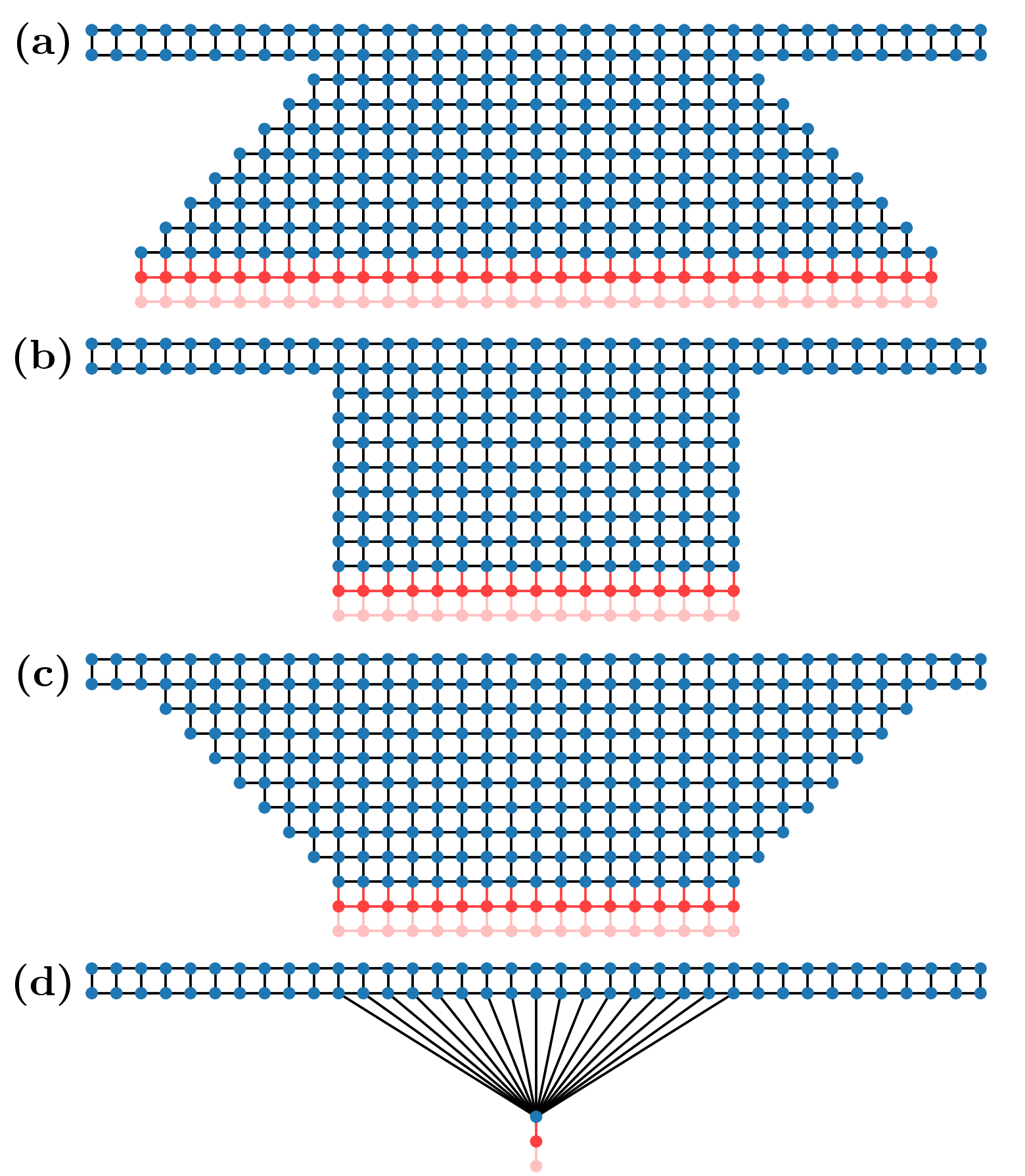}
    \caption{
    Four types of lead geometries defined on a finite difference grid: a) funnel, b) square, c) trumpet, d) point. 
    Bottom two rows in each image represent semi-infinite leads (indicated in red).
    }
    \label{fig:syst_types}
\end{figure}

Our calculations used a finite difference method to discretise the Hamiltonian given in Eq.~\ref{Eq:Hamiltonian}, resulting in a real space grid as shown in Fig.~\ref{fig:syst_types}.
Each vertex in the grid represents a discretisation site that has an associated on-site potential, and the lines represent hoppings that couple neighbouring sites.
Neighbouring sites in the grid are separated by a length of 3.125~nm.
This grid spacing was chosen by balancing computational efficiency and numerical accuracy, while ensuring that results converged.
Energies in our calculations are given in terms of a ``hopping energy'' $t$, defined as:
\begin{align}
\label{Eq:hopping}
    t &= \frac{\hbar^{2}}{2m^{*}_{}a^{2}}
\end{align}
where $a$ is the grid spacing.
The Peierls substitution modifies the hopping terms by a Peierls phase~\cite{Peierls1933a}, \textit{e.g.} if the hopping term between sites $r^{}_{1}$ and $r^{}_{2}$ is $t$, then:
\begin{equation}
    t \Rightarrow t\exp\left(\frac{e}{\hbar} \int^{r^{}_{2}}_{r^{}_{1}} \mathbf{A}(\mathbf{r}) \cdot d\mathbf{r}\right)
\end{equation}

We calculated our results using the software package KWANT, which uses a wave function matching method that is mathematically equivalent to the non-equilibrium Green's function method~\cite{Groth2014, Boumrar2020a}.
This model assumes ballistic transport (the electrons have mean free paths longer than the device length) and the zero temperature limit (no thermal broadening).
It is a fully quantum description that captures the wave-like nature of electrons as can be seen in the density plots in Fig.~\ref{fig:LDOS} where the trajectories are clearly not classical cyclotron trajectories.
We did not apply any bias between the leads in all of our calculations, so conductance was calculated using the Landauer-Büttiker formalism with zero-bias voltage, giving~\cite{Datta1995}:
\begin{align}
    \label{Eq:conductance}
    G &= \frac{2e^{2}}{h}\int T(E)\left[-\frac{\partial f_{0}(E)}{\partial E}\right]dE
\end{align}
where $h$ is Planck's constant, $T(E)$ is the transmission probability between the injector and collector, and $f_{0}(E)$ is the Fermi-Dirac distribution function.
In the zero temperature limit, the derivative of the Fermi-Dirac distribution function is a Dirac delta function, thus the integral in Eq.~\ref{Eq:conductance} is only evaluated at the exact energy chosen.
The conductance is then directly proportional to the transmission probability, with a proportionality constant given by the conductance quantum, $2e^{2}/{h}$.

The spin--projected probability densities in Fig.~\ref{fig:LDOS} are calculated by:
\begin{align}
    \rho^{}_{i,\pm} &= \left\langle \psi^{}_{i} \middle| \chi^{}_{\pm} \right\rangle \left\langle \chi^{}_{\pm} \middle| \psi^{}_{i} \right\rangle
    \label{Eq:Probability}
\end{align}
where for each site $i$, $\rho^{}_{i,\pm}$ is the spatially-resolved probability density for spin--up/down in $x$, $\psi^{}_{i}$ is the wavefunction element corresponding to that site, and $\chi^{}_{\pm}$ are the eigenvectors for the $x$ Pauli matrix.

\section{Results and discussion}
\subsection{Influence of device geometry}
\label{section:Shape}
An important consideration in TMF experiments is the influence of device geometry on the resulting signal.
Whilst the leads in actual devices are QPCs formed by smoothly varying potentials, here we model them as being confined by simple hard walls such that there are well-defined widths.
We present a more realistic model of the leads in section~\ref{section:QPCs}.
We define the width of a lead as the width of its narrowest section, and its shape by the position of that section relative to the contact point between the lead and the focusing area.

We find three different shapes used in TMF experiments in the literature.
First, there is the funnel-shaped lead (Fig.~\ref{fig:syst_types}a) which has its narrowest section at the contact point and becomes wider towards the electron source ~\cite{VanHouten1989,Potok2002,Yan2018}.
Second, the square-shaped lead (Fig.~\ref{fig:syst_types}b) maintains the same width throughout its entire length~\cite{Lo2017}.
Third, the trumpet-shaped lead (Fig.~\ref{fig:syst_types}c) is at its widest at the contact point and becomes narrower towards the electron source~\cite{Dedigama2006}. 
Figure~\ref{fig:syst_types}a--c shows the grids we used to simulate each of these shapes.
Finally, we also show a grid in Fig.~\ref{fig:syst_types}d that represents an idealised point (one-dimensional) lead, the end of which is connected to a number of sites in the focusing area, giving the lead an effective width.
For example, a 50~nm wide point lead refers to a 1D lead that is connected to 17 sites in the channel, which for a grid spacing of 3.125~nm is equivalent to a width of 50~nm (see Fig.~\ref{fig:syst_types}d).
\begin{figure}[th]
    \centering
    \includegraphics[width=0.95\linewidth]{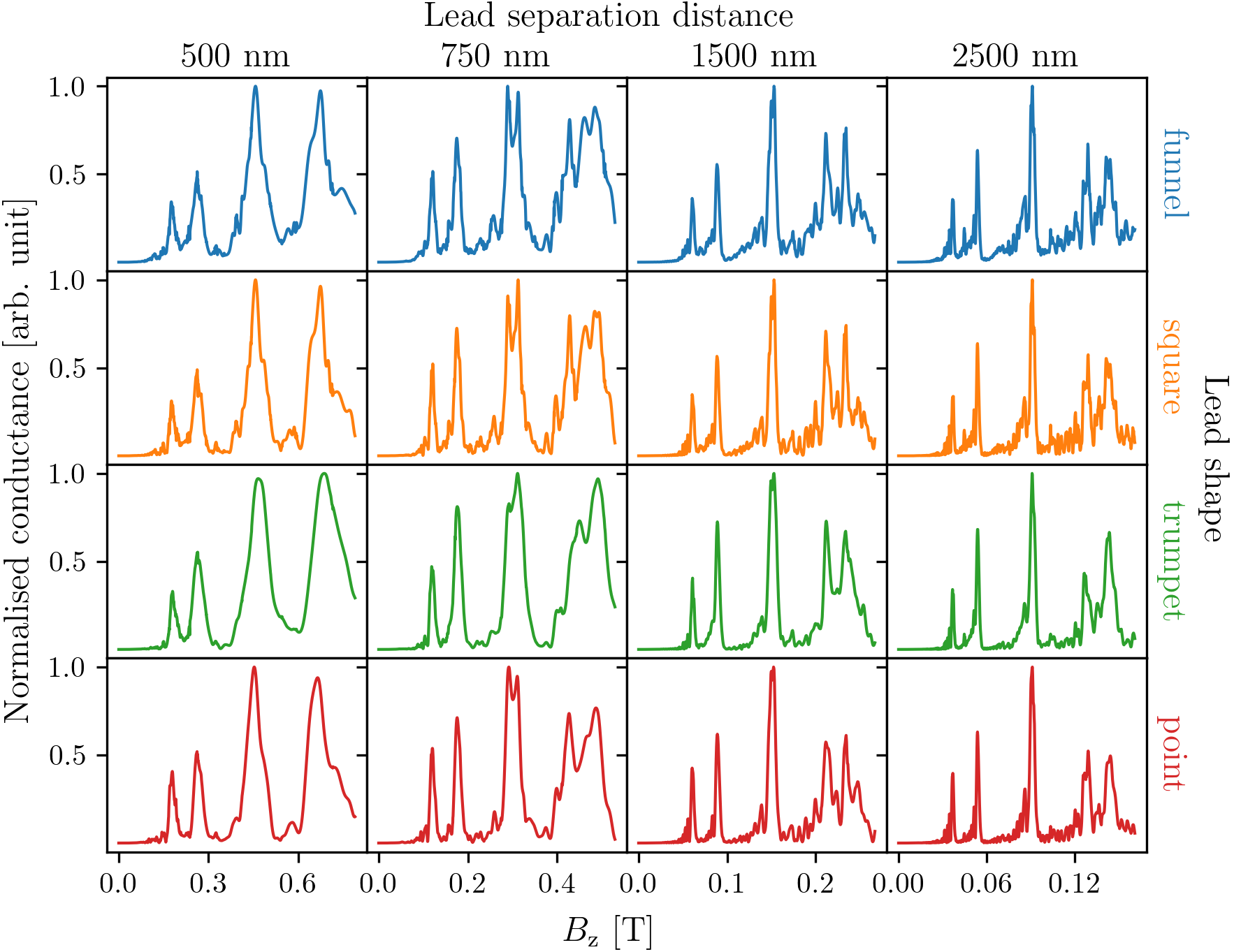}
    \caption{
    Conductance as a function of the out-of-plane magnetic field for a transverse magnetic focusing device with 50~nm wide point, trumpet, square, and funnel shaped leads with varying separation distances between the leads, calculated with the Fermi energy set to 7~meV.
    The conductance values are normalised to the maximum conductance in each plot.
    Note that plots in the same column share the same $x$-axis but different columns have differently scaled $x$-axes.
    }
    \label{fig:TvB_shape}
\end{figure}

\begin{figure}[ht]
    \includegraphics[width=0.95\linewidth]{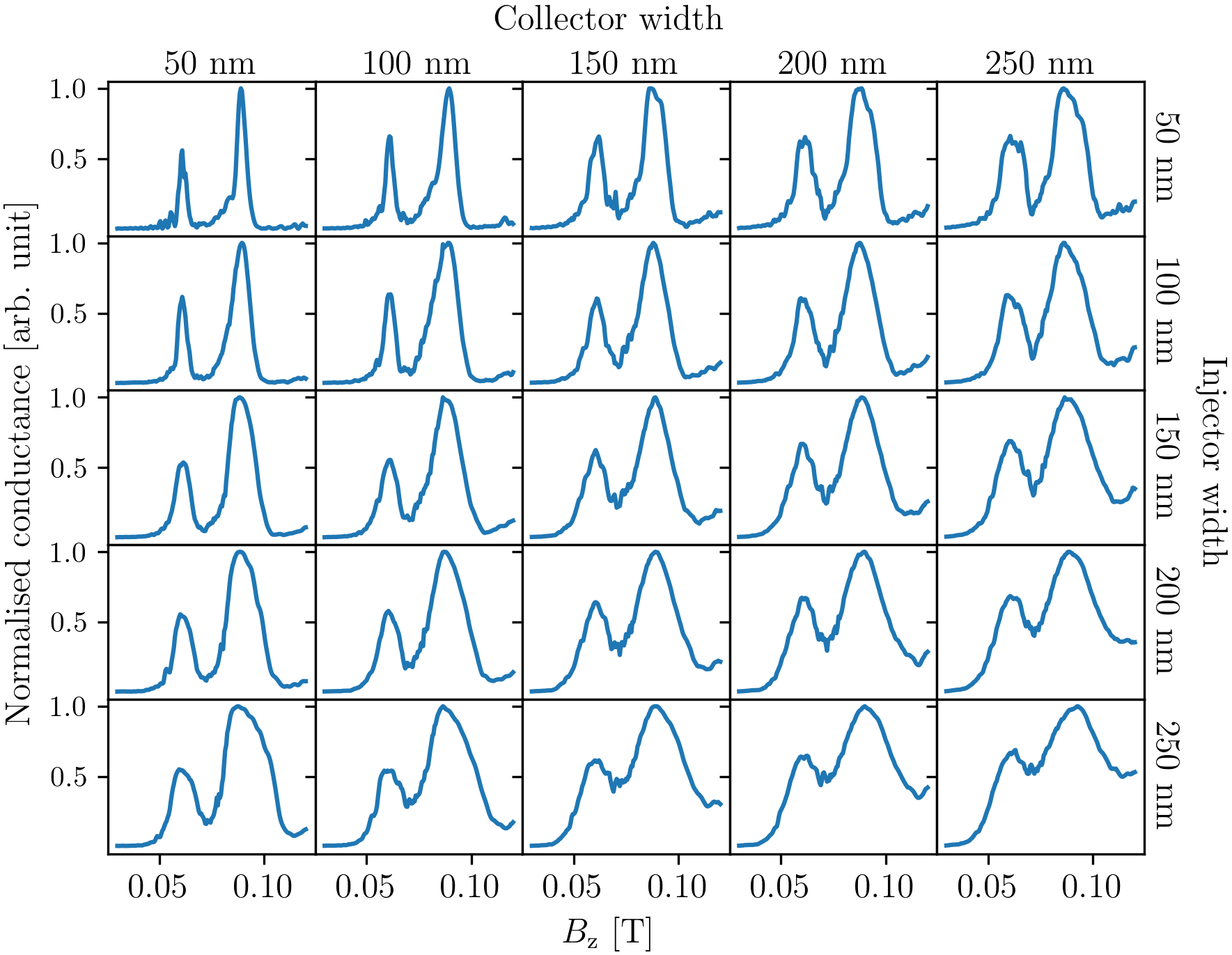}
    \caption{
    Conductance as a function of the out-of-plane magnetic field for transverse magnetic focusing devices with various combinations of injector and collector widths.
    All leads are trumpet shaped and set 1500~nm apart, with the Fermi energy set to 7~meV.
    Conductance values are normalised to the maximum value in each spectrum.
    }
    \label{fig:TvB_matrix}
\end{figure}

This idealised point lead will only have one transmission mode~\cite{Usaj2004}.
It is a simple theoretical model that ignores the effects of wave interference in the lead itself.
Similar grids for the point lead~\cite{Usaj2004} and square-shaped lead~\cite{Kormanyos2010} have been used in the literature for numerical simulations of TMF experiments.

The shape of the lead influences the impedance matching between the lead and the focusing area, and thus we expect that there is an effect on the measured conductance spectra.
The separation distance between the leads has also been shown to have an effect on the conductance spectrum~\cite{Rendell2015a}.
We calculated conductance as a function of the out-of-plane magnetic field for TMF devices using injector and collector leads of each of the aforementioned shapes with widths of 50~nm.
The separation distance between the leads was set to 500~nm, 700~nm, 1500~nm and 2500~nm.
Here a Fermi energy of $E_{\text{F}} = 7$~meV was used. 
Figure~\ref{fig:TvB_shape} shows the resulting conductance spectra with the conductance normalised to the maximum value in each data series.
Note that each column has a different scale for the $x$-axis since the peak positions shift towards lower magnetic field strengths as the lead separation is increased, which agrees with Eq.~\ref{Eq:larmor}.

We see that for each separation distance, the spectra for all lead shapes have features similar to the spectrum for the idealised lead (point).

The main peaks are easily resolvable in all the spectra, but finer structures such as the interference fringes are only clearly observed at the larger separations of 1500~nm and 2500~nm, when the peaks become sharper.
The shapes of the spectra change as the separation distance is changed, but stay relatively similar when the lead shape is changed for the same separation.
Interestingly, at a separation distance of 750~nm, a splitting appears in the second focusing peak, which should not have spin--splitting as discussed in Sec.~\ref{section:TMF}.
This splitting appears to be dependent on the separation distance as it is not noticeable at other separation distances.

The spectra for the trumpet shaped lead appear to have less fluctuations compared to the square and funnel shaped leads for all separation distances.
While the interference fringes before the first and second peaks are better resolved in the funnel and square shaped leads, we note that the magnitude of the fringes are comparable to the fluctuations seen in the peaks at higher magnetic field strengths.
Thus it would not be easy to differentiate between path interference fringes~\cite{Bladwell2017} and fluctuations from coincidental interference effects.
We chose to use the trumpet-shaped lead for subsequent calculations in this paper as it appeared to consistently give the best combination of low noise and clear resolution of features in the conductance spectrum, while still being an experimentally realistic geometry.

Another geometric aspect that affects the measured conductance spectra is lead width. 
Intuitively we would expect that the wider the leads, the wider the peaks would be in the spectra, since a wider injector would create a larger wavefront while a wider collector would collect more of that wavefront at each magnetic field strength. 
Figure~\ref{fig:TvB_matrix} shows the transmission spectra around the first peak pair in these spectra at $B^{}_{{z}} \approx 0.061$~T and $B^{}_{{z}} \approx 0.089$~T for TMF devices with the injector and collector widths set to combinations of 50~nm, 100~nm, 150~nm, 200~nm, and 250~nm.
Here a Fermi energy of 7~meV was used. 
We see that resolving the spin--split peaks depends on both the injector and collector widths.
The resolvability of the peaks is reduced as we move across each row and down each column of Fig.~\ref{fig:TvB_matrix}, with the clearest peaks observed when both leads are at their minimum widths (here 50~nm).

\begin{figure}[ht!]
    \includegraphics[width=0.95\linewidth]{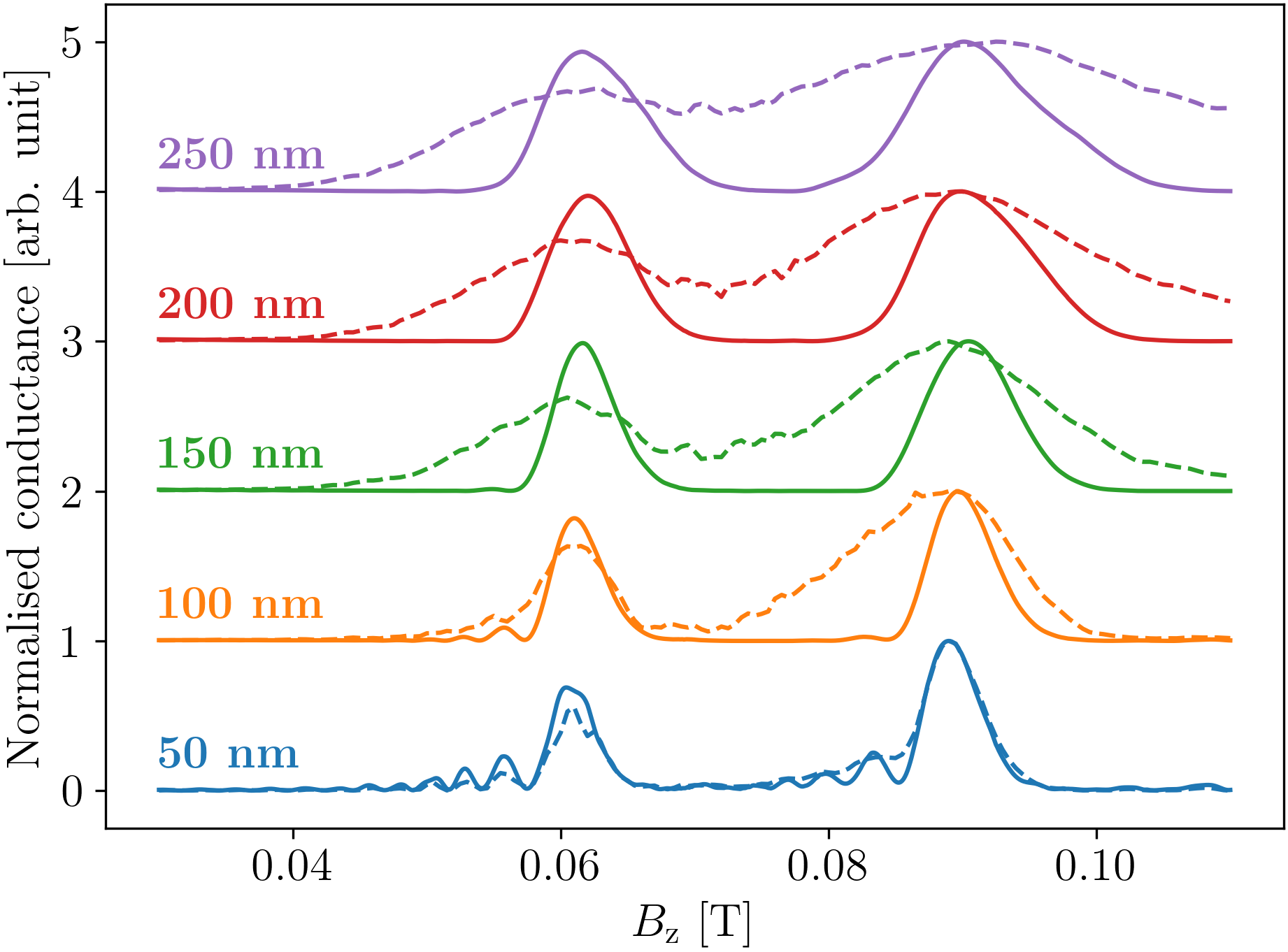}
    \caption{
    Conductance spectra calculated with the same parameters as in Fig.~\ref{fig:TvB_matrix} using idealised point leads (solid lines) and trumpet leads (dashed lines) with both injector and collector widths as labelled and set 1500 nm apart.
    Conductance values are normalised to the maximum value in each spectrum and offset for clarity.
    }
    \label{fig:TvB_width_point}
\end{figure}
\begin{figure}[ht]
    \includegraphics[width=0.95\linewidth]{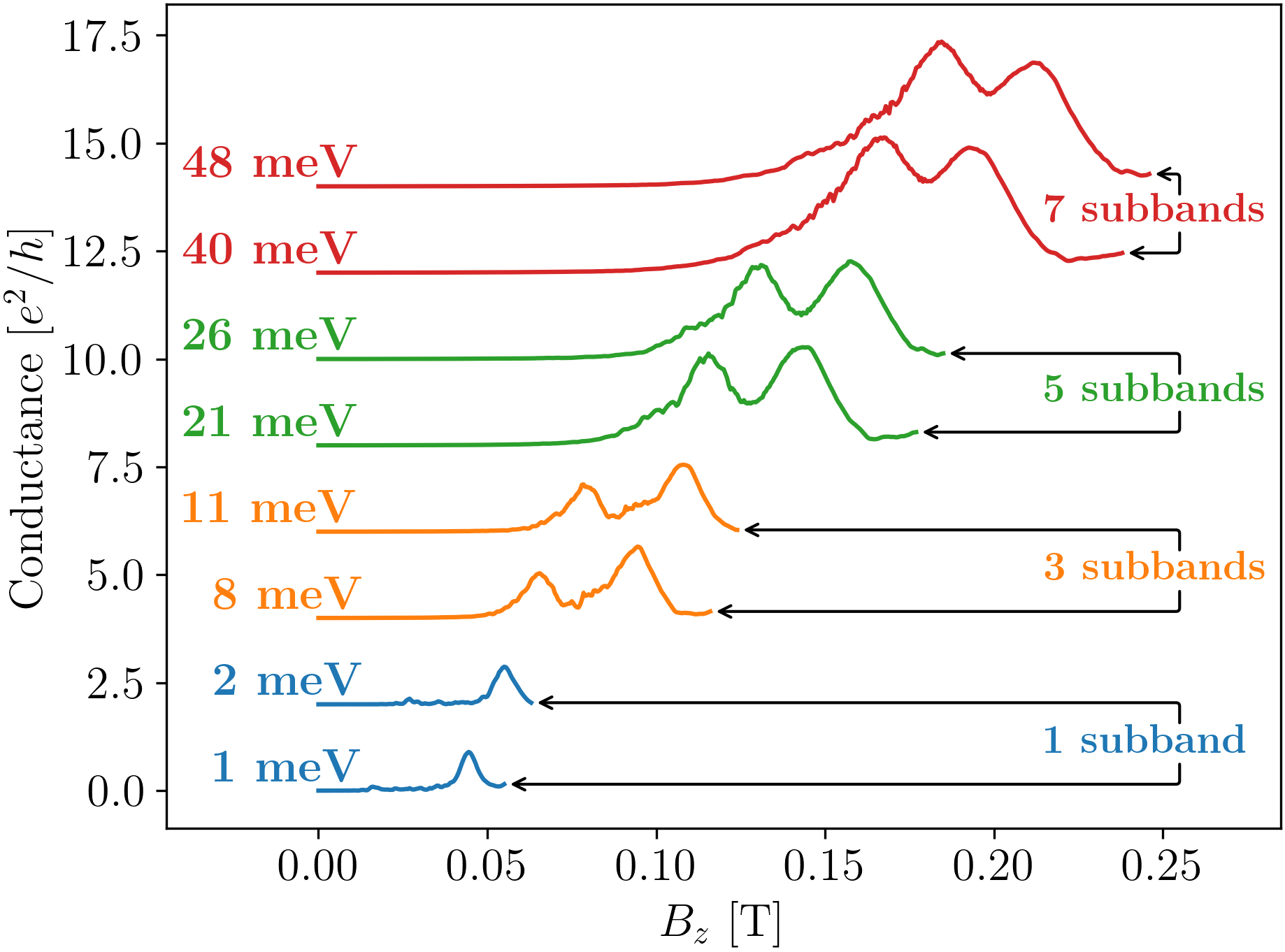}
    \caption{
    Conductance as a function of the out-of-plane magnetic field for a transverse magnetic focusing device with 100~nm wide trumpet shaped leads set 1500~nm apart, calculated with Fermi energies set between 8 and 75~meV.
    The Fermi energy crosses one (1~meV and 2~meV), three (8~meV and 11~meV), five (21~meV and 26~meV), and seven (40~meV and 48~meV) subbands as indicated, corresponding to the band structure shown in Fig.~\ref{fig:band_energies}.
    The shape of the peaks changes as more subbands are occupied, but stays the same for energies within the same subband.
    }
    \label{fig:TvB_energy}
\end{figure}

If we look at the conductance spectra of a TMF device using idealised leads of the same widths, as shown in Fig.~\ref{fig:TvB_width_point}, we can see a similar degradation in the resolvability of the peaks as the widths of the leads increase. 
The individual peaks broaden as the lead widens, and we observe that the interference fringes before the spin--split peaks are not observable at widths above 50~nm.
In general, the spin--split peaks broaden as the leads widen and the interference fringes become less observable.

\subsection{Subbands}
\label{section:Fermi}
Thus far we have not considered the number of subbands occupied by the electrons when we choose the Fermi energy in our calculations. 
In experiments this can be controlled by changing the gate voltage at the leads to have conductance values that correspond to the number of subbands occupied~\cite{Chuang2015}.
Adjusting the gate voltage in experiments effectively shifts the band structure within the leads, whilst the Fermi energy stays constant.

\begin{figure}[ht]
    \includegraphics[width=0.95\linewidth]{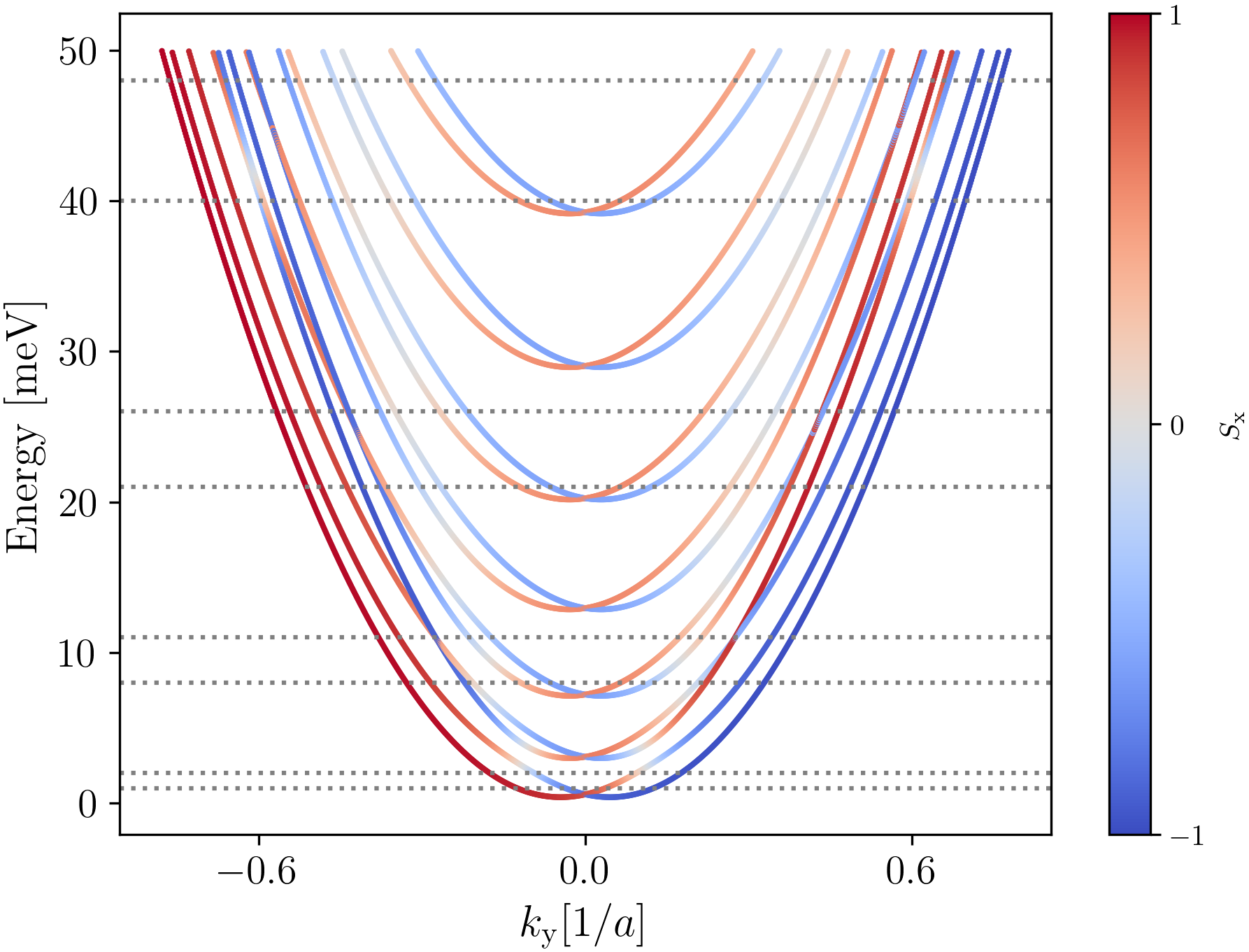}
    \caption{
    1D band structure for a 100~nm wide infinite wire with Rashba spin--orbit interaction present.
    The colour bar indicates the $x$-component of the electron spin, with spin--up shown as red and spin--down shown as blue.
    The horizontal dotted lines indicate energies used for calculating the conductance plots in Fig.~\ref{fig:TvB_energy}.
    }
    \label{fig:band_energies}
\end{figure}

In our calculations, we chose to shift the Fermi energy instead for numerical simplicity.
While this will shift the positions of the peaks in the conductance spectra for different Fermi energies, the effect this has on the overall features of the spectra (\textit{e.g.}~peak shapes) is equivalent to shifting the band structure.
We will present a more realistic model of the leads as QPCs controlled by gate voltages in Sec.~\ref{section:QPCs}.
We examine in Fig.~\ref{fig:TvB_energy} how the features of the peaks at the first focusing magnetic field strength change as the the Fermi energy shifts through the lowest four odd numbers of subbands.
The TMF device simulated has trumpet-shaped leads with widths of 100~nm separated by 1500~nm.
To interpret these results, Fig.~\ref{fig:band_energies} shows the 1D band structure calculated for a nanowire with a width of 100~nm and the dotted lines indicate the corresponding Fermi energies shown in Fig.~\ref{fig:TvB_energy}.

In Fig.~\ref{fig:TvB_energy}, we see that the amplitude of the peaks increases as the Fermi energy increases, as more conducting modes are available when more subbands are occupied.
The ratio of the first peak amplitude to the second peak amplitude changes from low:high to high:low as more subbands are occupied, but the peak shapes and amplitudes do not change significantly when the Fermi energy changes within the same subband.
In the spectra that correspond to one occupied subband (blue), the first spin--split peak is greatly suppressed.
This strong spin polarisation is explained by adiabatic transitions between anti-crossed subbands as the lead widens~\cite{Eto2005}.
However, note that we do not observe the suppression of the first spin--split peak in the spectra along the first row and column of Fig.~\ref{fig:TvB_matrix}, even though there is only one occupied subband at 7~meV for a 50~nm wide nanowire.
For a 50~nm wide nanowire, the anti-crossing between the first and second subbands occurs at an energy close to the minimum of the third subband, a much higher energy compared to the case for a 100~nm wide nanowire (see Fig.~\ref{fig:band_energies}).
This suggests the 50~nm wide lead does not widen enough for the electron to transition through the anti-crossing point, thus the first spin--split peak is not suppressed.

\begin{figure}[ht]
    \includegraphics[width=0.95\linewidth]{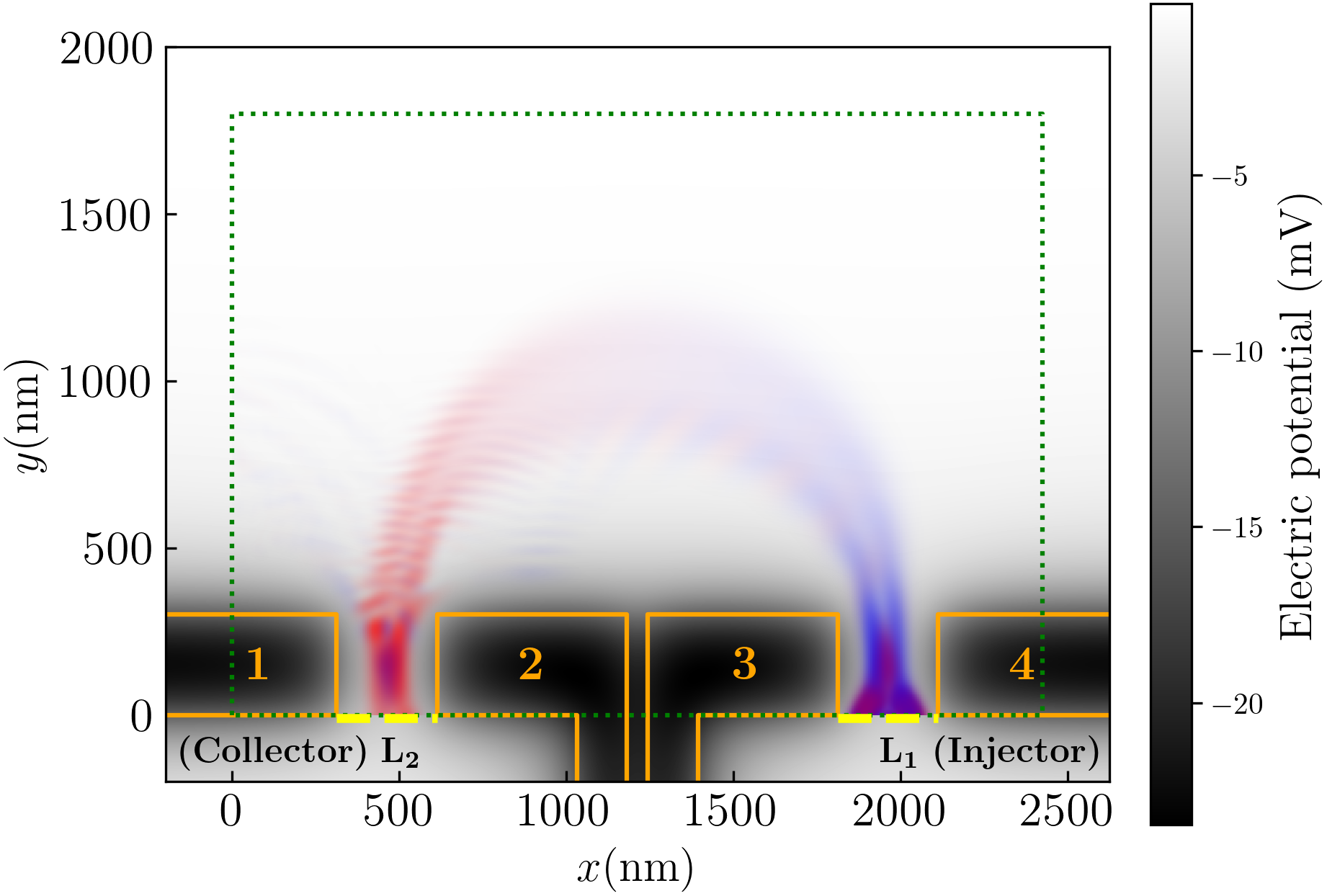}
    \caption{
    Calculated potential field in 2DEG from 40~mV potential applied at surface gates labelled 1, 2, 3 and 4 (marked by orange solid lines).
    The green dotted box denotes the area simulated in calculations. 
    The yellow dashed lines indicate where 300~nm wide semi-infinite leads were attached.
    The overlaid spin--projected density plot shows the probability density of an electron with a Fermi energy of 7~meV injected at L$_{1}$ and collected at L$_{2}$ under an out-of-plane magnetic field of 0.0957~T.
    The density plot has the same colour scheme as in Fig.~\ref{fig:LDOS}.
    }
    \label{fig:Potential}
\end{figure}

The band structure significantly affects the features of the TMF spectra, therefore it is vitally important to ensure that the number of subbands occupied and the spin compositions of the subbands are known when interpreting TMF spectra.

\subsection{Simulation of realistic QPCs}
\label{section:QPCs}
In experiments, the injector and collector leads in Fig.~\ref{fig:TMF_system}a are QPCs formed by applying voltage to surface gates, depleting electrons within the 2DEG under the gates~\cite{VanHouten1989}.
Changing the gate voltages effectively changes the width of the QPC~\cite{VanHouten1989} and shifts the band structure within the QPC.
We can approximate the potential field within the 2DEG applied by the surface gates using the method derived by Davies \textit{et al.}~\cite{Davies1995}.
Adding this potential field as a position dependent potential term to the Hamiltonian in Eq.~\ref{Eq:Hamiltonian} allows us to model the TMF device more realistically.

For our calculations we used the potential generated from surface gates with shapes illustrated by the orange lines in Fig.~\ref{fig:Potential}.
The injector QPC is formed by gates 3 and 4, while the collector QPC is formed by gates 1 and 2.
The gates forming each QPC are separated by 300~nm.
To avoid asymmetric biasing in the QPCs, we apply equal voltages on gates 1 and 2, as well as gates 3 and 4. 
We define the injector and collector gate voltages ($V_{\mathrm{g}}$) as the voltage applied to the gates forming the respective QPCs.
The semi-infinite leads attached below each QPC have the same potential profile as the sites adjacent to the semi-infinite leads.

\begin{figure}[ht]
    \includegraphics[width=0.95\linewidth]{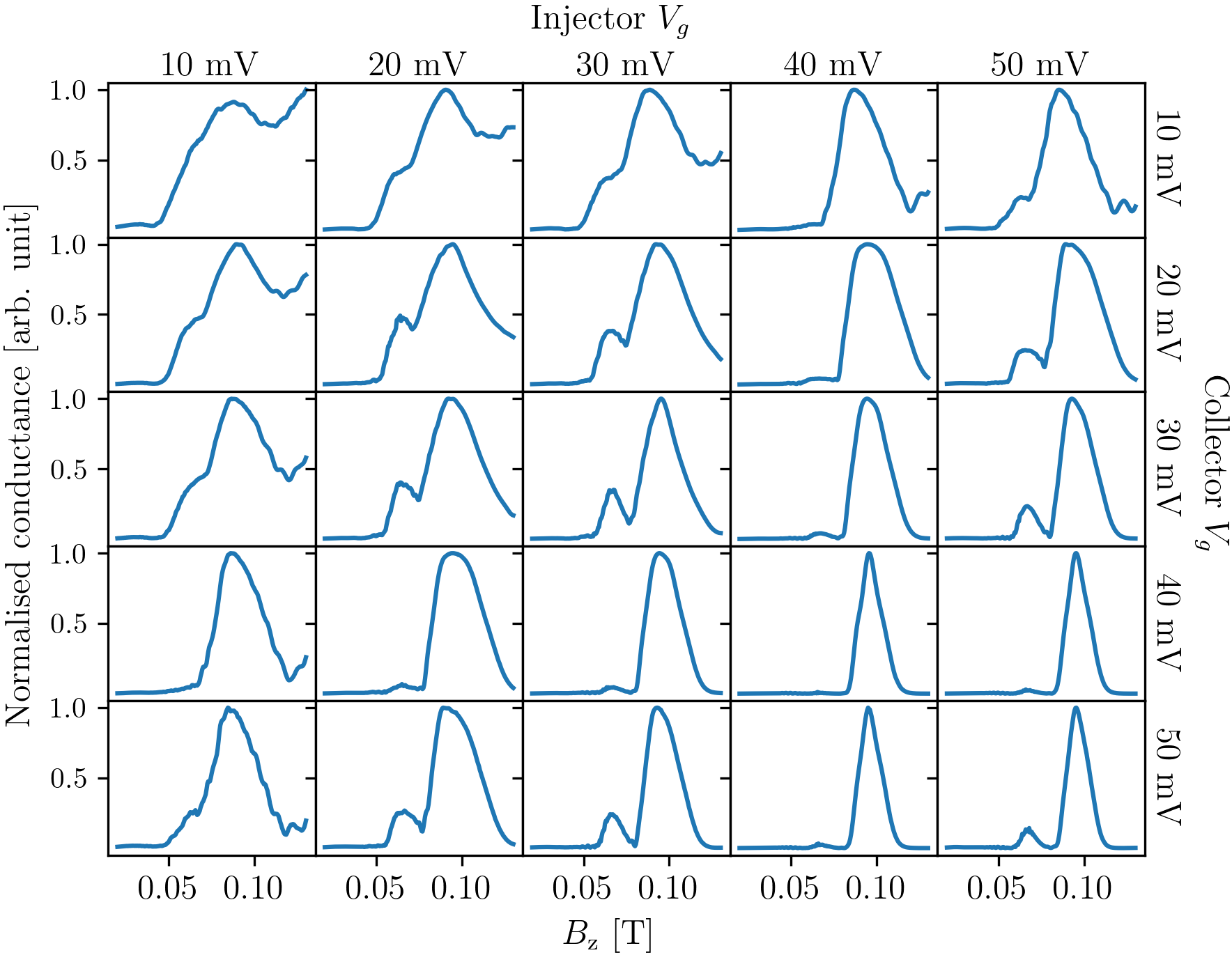}
    \caption{
    Conductance as a function of the out-of-plane magnetic field for transverse magnetic focusing devices with various combinations of gate voltages at the injector and collector.
    All leads are set 1500~nm apart, with the Fermi energy set to 7~meV.
    Conductance values are normalised to the maximum value in each spectrum.
    }
    \label{fig:QPC_matrix}
\end{figure}

We calculated the transmission spectra for a TMF device with a lead separation of 1500~nm and a Fermi energy of 7~meV.
The injector and collector gate voltages were set to combinations of 10~mV, 20~mV, 30~mV, 40~mV, and 50~mV (Fig.~\ref{fig:QPC_matrix}).
The conductance values in Fig.~\ref{fig:QPC_matrix} are normalised for better comparison of the peak features.
However, we note that the conductance values for gate voltages of 50~mV (bottom row and right-most column in Fig.~\ref{fig:QPC_matrix}) are actually several orders of magnitude smaller than the conductance values in the other spectra.
Also, the conductance values for gate voltages of 40~mV are close to one conductance quanta.
This suggests that the QPC formed by gate voltages of 40~mV is only allowing transmission in the lowest subband, and at 50~mV, the only transmission occurring is from tunneling electrons. 
We see that similar to Fig.~\ref{fig:TvB_matrix}, the peaks broaden as the effective width of the QPCs increase (\textit{i.e.}~the voltage decreases).
However, unlike Fig.~\ref{fig:TvB_matrix}, reducing the effective width of the QPC does not allow the interference fringes to be resolved even with the narrowest QPCs.
This suggests that to observe the interference fringes experimentally, the QPCs will need to have sharp hard wall-like potential profiles.

We also see that the first peak is greatly suppressed at gate voltages of 40~mV, similar to the single subband spectra in Fig.~\ref{fig:TvB_energy}.
Again, this spin polarisation can be explained by adiabatic transitions between anti-crossed subbands as the QPC narrows and widens~\cite{Eto2005}.
Thus, we can see that the effect of changing the gate voltage is a combination of changing the width and Fermi energy in our previous simpler model.
While the hard wall model of leads is less realistic, it allows us to explore the effects of width and Fermi energy independently.

\subsection{Disorder}
\label{section:Disorder}
Devices used in experiments are usually not perfect, they exhibit defects, impurities, surface roughness and other imperfections. 
This causes electrons to scatter as they travel through the device. 
Here we simulate disorder using an Anderson localisation model~\cite{Anderson1958}, where we add a random onsite potential with values chosen from a uniform distribution within the range $[-U/2,U/2]$ to each site within the focusing area, but not within the semi-infinite leads (the red sites in Fig.~\ref{fig:syst_types}).
$U$ is defined as the disorder strength and can be related to the mean free path ($l^{}_{\mathrm{mfp}}$) of the electrons using Fermi's golden rule~\cite{ihn2009} (Derivation shown in supplementary information):
\begin{align}
    l^{}_{\mathrm{mfp}}=48a\frac{\sqrt{E^{}_{\mathrm{F}}/t}}{(U/t)^2}
    \label{Eq:mfp}
\end{align}
where $a$ is the length of the grid spacing, $E^{}_{\mathrm{F}}$ is the Fermi energy, and $t$ is the hopping energy as defined in Eq.~\ref{Eq:hopping}.
\begin{figure}[ht]
    \includegraphics[width=0.95\linewidth]{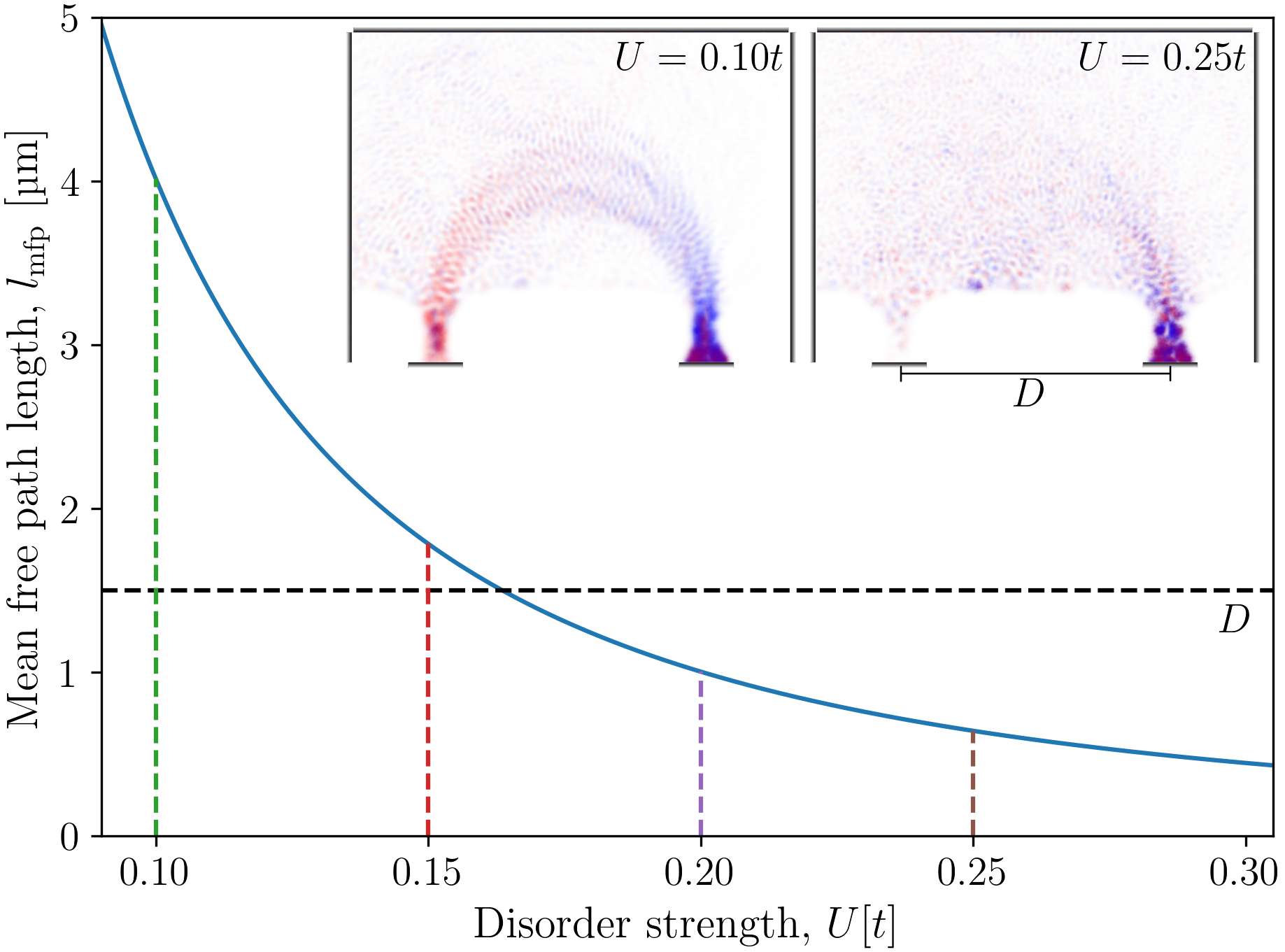}
    \caption{
    Mean free path length ($l^{}_{\mathrm{mfp}}$) of an electron in a device with disorder strength $U$ calculated using Eq.~\ref{Eq:mfp}.
    The lattice constant ($a$) and Fermi energy ($E^{}_{\mathrm{F}}$) were set to be 3.125~nm and 7~meV respectively.
    The inset plots show the probability density for devices at a magnetic field strength of $B^{}_{{z}}=0.089$~T, with particular realisations of disorder strengths 0.10$t$ and 0.25$t$.
    The horizontal black dashed line at $l^{}_{\mathrm{mfp}}=1500$~nm indicates the distance $D$ between the injector and collector leads in the device used in Fig.~\ref{fig:Disorder}.
    The vertical dashed lines correspond to the disorder strengths used in Fig.~\ref{fig:Disorder}.
    }
    \label{fig:disorder_mfp}
\end{figure}
For our calculations, we set $a$ to be 3.125~nm and $E^{}_{\mathrm{F}}$ to be 7~meV.
The mean free path lengths at varying disorder strengths calculated from Eq.~\ref{Eq:mfp} is shown in Fig.~\ref{fig:disorder_mfp}.
The two inset probability density plots show that the spin--split trajectories become less well-defined as the mean free path length becomes shorter than the distance between the injector and collector leads.

\begin{figure}[ht]
    \includegraphics[width=0.95\linewidth]{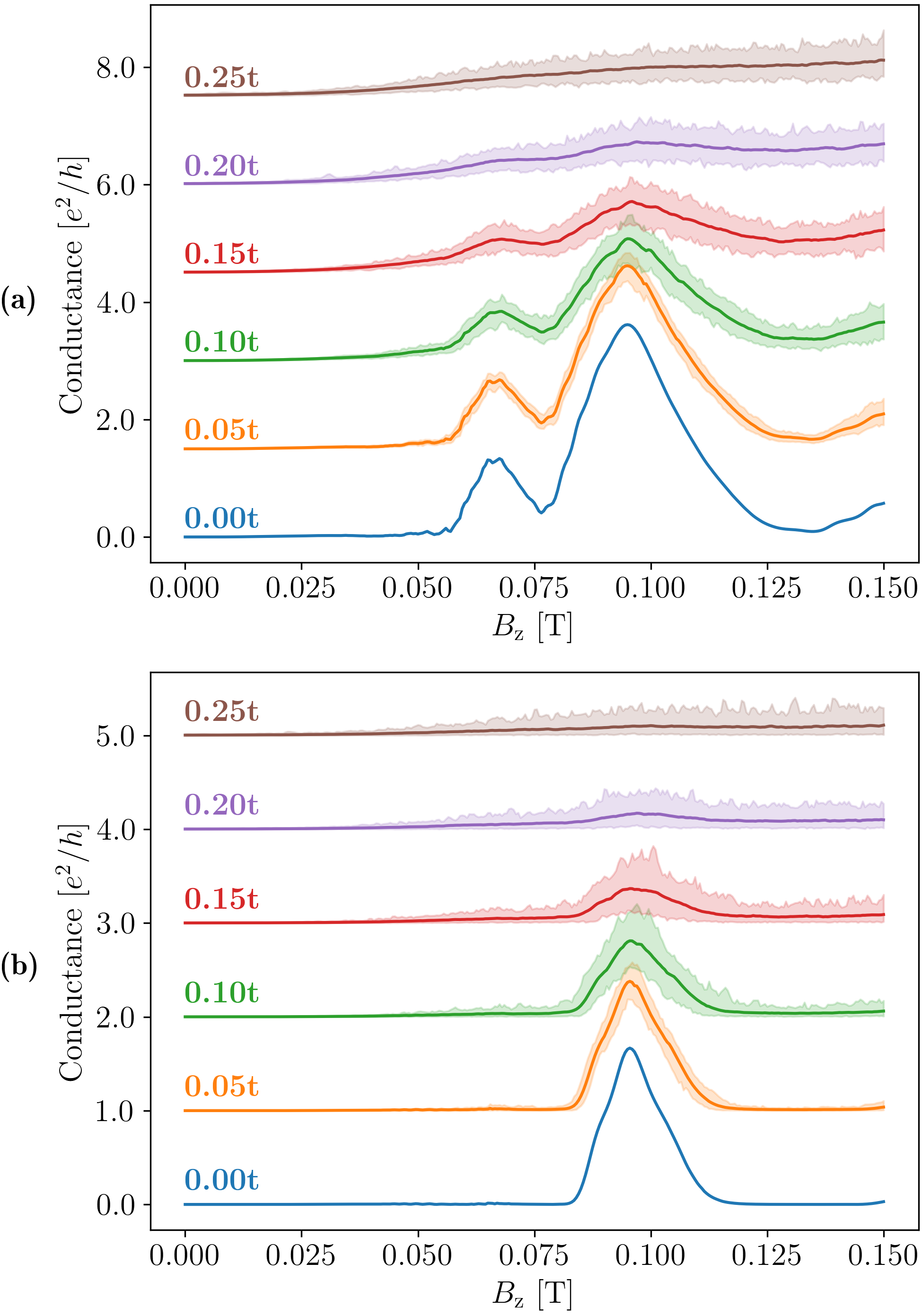}
    \caption{
    Conductance as a function of the out-of-plane magnetic field for a transverse magnetic focusing device with injector and collector QPCs set 1500~nm apart, and both gate voltages set to (a) 30~mV and (b) 40~mV.
    Fermi energies were set to 7~meV.
    Disorder was implemented in the systems by applying onsite potentials chosen at random from a uniform distribution within the range $[-U/2,U/2]$.
    Solid lines show the mean values for 201 realisations of that particular disorder strength, and the corresponding shaded areas show the peak-to-peak variance.
    }
    \label{fig:Disorder}
\end{figure}

Figure~\ref{fig:Disorder} shows the conductance spectra calculated for a TMF device for various disorder strengths $U$. 
We performed conductance calculations for 201 different disorder realisations at each disorder strength.
The mean of the conductance values at each magnetic field strength and disorder strength are shown as solid lines in Fig.~\ref{fig:Disorder}, and the shaded areas show the variance of the conductance values for each disorder strength.
We observe that as disorder increases, the amplitude of the conductance peaks decrease as expected since scattering reduces the probability that the electron reaches the collector lead.
The spin--split peaks at $B^{}_{{z}}\approx0.061$~T and $B^{}_{{z}}\approx0.089$~T are greatly suppressed and indistinguishable from signal noise at disorder strengths of $0.25t$ and above.
This agrees with our calculation that the mean free path of the electron at $U=0.25t$ is less than half the distance between the injector and collector leads (Fig.~\ref{fig:disorder_mfp}), thus the electron is likely to be scattered without reaching the collector.

We only simulated disorder within the weak localisation regime where the mean free path length is much greater than the Fermi wavelength.
However, even within this regime, the characteristic features in the conductance spectra rapidly became indistinguishable as the devices were increasingly disordered.
Therefore, this implies that to observe the finer structure in the spectra such as the interference fringes, the device used has to be relatively free from disorder.

\section{Conclusions}
\label{section:Conclusion}
We have shown the geometry of the device used in a TMF experiment plays a significant role in the measured characteristics of the conductance spectra.
The shape of the leads used does not significantly affect the separation of the spin--split peaks, but does affect the level of noise or fluctuations in the spectra.
The separation distance between the injector and collector leads can significantly alter the features of the spectra.
The width of the leads used significantly affects the resolution of the finer features in the TMF spectra, \textit{e.g.}~the interference fringes.
Furthermore, the spin composition of the subbands and the number of subbands occupied both affect the ratio of the spin--split peak amplitudes.
Thus, accurate knowledge of the device geometry and band structure is essential when interpreting TMF results.

As a result of these effects, the potentials forming QPCs in a TMF device need to be very well-defined in order to observe finer structures in the TMF spectra, \textit{e.g.}~the interference fringes.
In addition, we show that the TMF device is sensitive to disorder and features in the measured conductance spectrum quickly become obfuscated as the device becomes disordered.
The use of real-space numerical simulation of TMF experiments allows us to disentangle geometric effects and device imperfections from the spin--split structure and the interference fringes.
This in turn allows for unambiguous interpretation of TMF experiments in the future.

\section*{Abbreviations}
TMF: Transverse magnetic focusing; 2DEG: Two-dimensional electron gas; 2DHG: Two-dimensional hole gas; QPC: Quantum point contact

\section*{Declarations}
\subsection*{Availability of data and materials}
The datasets used and/or analyzed during the current study are available from the corresponding author on reasonable request.

\subsection*{Competing interests}
The authors declare that they have no competing interests.

\subsection*{Funding}
This work was funded in part by the Australian Research Council Centre of Excellence in Future Low-Energy Electronics Technologies (project number CE170100039).
Computational resources and services were provided by the National Computational Infrastructure (NCI), which is supported by the Australian Government.

\subsection*{Authors' contributions}
YKL wrote the code, performed the calculations and drafted the manuscript. All authors analysed and interpreted the results, and revised the manuscript. All authors read and approved the final manuscript.

\subsection*{Acknowledgements}
We acknowledge useful conversations with S. Bladwell, O. Sushkov and D. Culcer. 

\bibliography{TMF_paper}

\onecolumngrid
\section{Supplementary Information: Derivation of mean free path formula (Eq. 11)}
\label{A:derivation_of_mfp}
Here we derive Eq. 11.
We begin with an equation for the scattering time in the single subband case ($\tau_{0}$) given in pg. 168 of \cite{ihn2009}:
\begin{align}
    \frac{\hbar}{\tau_{0}} &= n_{i}\frac{m^{*}}{2\pi\hbar^{2}}\int^{2\pi}_{0}d\varphi\left\langle\left|v^{(i)}(\vec{q})\right|^{2}\right\rangle_{\mathrm{imp}}\left(1-\cos\varphi\right) \label{Eq:scattering_time}
\end{align}
where $n_{i}$ is the areal density of scatterers, $\varphi$ is the scattering angle, and $\left\langle\left|v^{(i)}(\vec{q})\right|^{2}\right\rangle_{\mathrm{imp}}$ is the ensemble average of the squared scattering potential matrix elements.

The areal density of scatterers is defined as:
\begin{align}
    n_{i} &:= \frac{N_{i}}{A}
\end{align}
where $N_{i}$ is the number of scattering centers within the normalization area $A$.

In our disorder simulations we applied an onsite potential to every discretisation site in the device, which act as the scatterers.
Since there is always one scatterer within each unit cell in the discretisation grid, our normalization area is the area of one unit cell, $a^{2}$, where $a$ is the grid spacing length.
Therefore our areal density of scatterers is one per unit cell area:
\begin{align}
    n_{i} &= \frac{1}{a^{2}} \label{Eq:n_i}
\end{align}

The ensemble average of the squared scattering potential matrix elements can be related to the ensemble average of the squared scattering matrix elements by~\cite{ihn2009}:
\begin{align}
    \left\langle\left|v^{(i)}(\vec{q})\right|^{2}\right\rangle_{\mathrm{imp}} &= \frac{A^{2}}{N_{i}}\left\langle\left|\left\langle m\mathbf{k}_{m}\left|V\right|n\mathbf{k}_n\right\rangle\right|^{2}\right\rangle_{\mathrm{imp}}
    \label{Eq:scattering_potential}
\end{align}
where $V$ is the scattering potential.
Since our scattering potential is a uniform random distribution within the range $\left[-\frac{U}{2},\frac{U}{2}\right]$, independent of $\mathbf{k}$, $\left\langle m\mathbf{k}_{m}\left|V\right|n\mathbf{k}_n\right\rangle=V$. The probability distribution of $V$, $f_{{x}}(V)$ is given by:
\begin{align}
    f_{{x}}(V) &= \frac{1}{V_{max}-V_{min}} \nonumber\\
    &= \left[\frac{U}{2}-\left(-\frac{U}{2}\right)\right]^{-1} \nonumber\\
    &= \frac{1}{U}
\end{align}
Applying the law of the unconcious statistician to Eq.~\ref{Eq:scattering_potential}, we obtain:
\begin{align}
    \left\langle\left|v^{(i)}(\vec{q})\right|^{2}\right\rangle_{\mathrm{imp}} &= \frac{A^{2}}{N_{i}}\left\langle\left|V\right|^{2}\right\rangle_{\mathrm{imp}} \nonumber\\
    &= \frac{A}{n_{i}}\int^{\frac{U}{2}}_{-\frac{U}{2}}V^{2}f_{{x}}(V)dV \nonumber\\
    &= \frac{a^{2}}{n_{i}}\frac{U^{2}}{12} \label{Eq:V_2}
\end{align}

We can rewrite the scattering time in terms of the mean free path length $l_{\mathrm{mfp}}$ and Fermi velocity $v_{\mathrm{F}}$:
\begin{align}
    \tau_{0} &= \frac{l_{\mathrm{mfp}}}{v_{\mathrm{F}}} \nonumber\\
     &= l_{\mathrm{mfp}}\sqrt{\frac{m^{*}}{2E_{\mathrm{F}}}} \label{Eq:tau}
\end{align}
Substitute Eqs.~\ref{Eq:n_i},~\ref{Eq:V_2} and~\ref{Eq:tau} into Eq.~\ref{Eq:scattering_time}:
\begin{align}
    \frac{\hbar}{\tau_{0}} &= n_{i}\frac{m^{*}}{2\pi\hbar^{2}}\int^{2\pi}_{0}d\varphi\left\langle\left|v^{(i)}(\vec{q})\right|^{2}\right\rangle_{\mathrm{imp}} \left(1-\cos\varphi\right) \nonumber\\
    \frac{\hbar}{l_{\mathrm{mfp}}}\sqrt{\frac{2E_{\mathrm{F}}}{m^{*}}} &= n_{i}\frac{m^{*}}{2\pi\hbar^{2}}\frac{a^{2}}{n_{i}}\frac{U^{2}}{12}{2\pi} \nonumber\\
    l_{\mathrm{mfp}} &= \sqrt{\frac{2}{m^{*}}} \frac{12\hbar^{3}}{m^{*}a^{2}} \frac{\sqrt{E_{\mathrm{F}}}}{U^{2}} \nonumber
\end{align}
We want to express the energy terms as fractions of the hopping energy $t$.
\begin{align}
    l_{\mathrm{mfp}} &= \sqrt{\frac{2}{m^{*}}} \frac{12\hbar^{3}}{m^{*}a^{2}} \frac{\sqrt{E_{\mathrm{F}}/t}}{U^{2}/t^{2}} \frac{\sqrt{t}}{t^{2}} \nonumber\\
    &= 48a\frac{\sqrt{E_{\mathrm{F}}/t}}{U^{2}/t^{2}}
\end{align}

\end{document}